\documentclass{Interspeech2024}
\usepackage{amsmath,graphicx,xcolor,lipsum}

\newcommand\blfootnote[1]{%
  \begingroup
  \renewcommand\thefootnote{}\footnote{#1}%
  \addtocounter{footnote}{-3}%
  \endgroup
}
\def\L{{\cal L}}
\interspeechcameraready 
\title{Non-Intrusive Speech Intelligibility Prediction for Hearing Aids using Whisper and Metadata}

\name{Ryandhimas E. Zezario$^1$$^2$, Fei Chen$^3$, Chiou-Shann Fuh$^1$, Hsin-Min Wang$^2$, Yu Tsao$^2$}
\address{
  $^1$National Taiwan University,
  $^2$Academia Sinica,\\
  $^3$Southern University of Science and Technology}
\email{\{ryandhimas, yu.tsao\}@citi.sinica.edu.tw}
\keywords{speech intelligibility, hearing aid, hearing loss, weak supervision, cross-domain features}

\begin{document}

\maketitle

\begin{abstract}
Automated speech intelligibility assessment is pivotal for hearing aid (HA) development. In this paper, we present three novel methods to improve intelligibility prediction accuracy and introduce MBI-Net+, an enhanced version of MBI-Net, the top-performing system in the 1st Clarity Prediction Challenge. MBI-Net+ leverages Whisper's embeddings to create cross-domain acoustic features and includes metadata from speech signals by using a classifier that distinguishes different enhancement methods. Furthermore, MBI-Net+ integrates the hearing-aid speech perception index (HASPI) as a supplementary metric into the objective function to further boost prediction performance. Experimental results demonstrate that MBI-Net+ surpasses several intrusive baseline systems and MBI-Net on the Clarity Prediction Challenge 2023 dataset, validating the effectiveness of incorporating Whisper embeddings, speech metadata, and related complementary metrics to improve prediction performance for HA.

%\noindent\textbf{Index Terms}: speech intelligibility, hearing aid, hearing loss, weak supervision, cross-domain features

\end{abstract}
\blfootnote{This work was supported by the National Natural Science Foundation of China (Grant Nos. 62371217).}
\section{Introduction}
Accurate metrics for predicting speech intelligibility are crucial for optimizing the efficacy of various speech-related applications, such as speech enhancement \cite{loizou2007speech, 10248166}, hearing aid (HA) devices \cite {barker22_interspeech, 10362991}, and telecommunications \cite {yi22b_interspeech, POLQA}. The most direct method of evaluation is to conduct a human listening test. However, for a fair evaluation of results, a large-scale listening test is typically required, which can be costly and less practical. Therefore, various objective speech intelligibility measures have been proposed. Traditionally, these measures are derived based on signal processing and psychoacoustic knowledge, such as speech intelligibility index (SII) \cite{ref_36}, extended SII (ESII) \cite{ref_37}, speech transmission index (STI) \cite{ref_38}, short-time objective intelligibility (STOI) \cite{ref_39}, modified binaural short-time objective intelligibility (MBSTOI) \cite{ANDERSEN20181}, gammachirp envelope
similarity index (GESI) \cite{irino22_interspeech}, and the hearing aid speech perception index (HASPI)\cite{katehaspi}. However, these traditional measures have limited applicability, as they generally require a clean reference, which may not always be available in real-world scenarios.

With the emergence of deep learning models, numerous non-intrusive automatic speech assessment models have been developed, such as \cite{ref_56, mbnet_mos, ssl-mos, yang22o_interspeech, saeki22c_interspeech, mosa-net, zezario2022mti, Cooper2024e24}. Meanwhile, many studies have focused on developing speech intelligibility prediction models for HA \cite{chiang2023multiobjective,kates2014hearingb,tu22_interspeech, edozezario22_interspeech, MAWALIM2023109663}. For example, \cite{chiang2023multiobjective} encodes the hearing loss pattern into a vector, merges it with speech signals, and feeds it into a deep learning model to predict two hearing aid evaluation metrics: HASPI and the hearing aid speech quality index (HASQI) \cite{kates2014hearingb}. Besides, \cite{tu22_interspeech} utilizes hidden layer representations of an automatic speech recognition (ASR) model as acoustic features for predicting speech intelligibility scores. Furthermore, \cite{edozezario22_interspeech} introduces a multi-branched speech intelligibility prediction model, namely MBI-Net, consisting of two branch modules to estimate the frame-level scores of the left and right channel inputs; the outputs of the two branches are then concatenated and fused in a linear layer to produce the final intelligibility prediction score for HA.

While MBI-Net \cite{edozezario22_interspeech} demonstrates strong performance and achieves top results in the 1st Clarity Prediction Challenge \cite{barker22_interspeech}, there are several options available to further enhance its performance. In this study, we explore three methods and introduce an advanced version of MBI-Net, termed MBI-Net+. First, MBI-Net+ incorporates Whisper \cite{Whisper} embeddings to generate cross-domain features. Considering Whisper underwent training using a comprehensive dataset comprising 680,000 utterances along with their respective transcripts, we assume that the extracted embedding features possess the capability to encapsulate rich phonetic information. Second, MBI-Net+ includes metadata from speech signals to mitigate potential prediction bias. This addition involves incorporating an extra classifier to distinguish different types of speech enhancement methods (front-end processors in HAs). We assume that if the model can optimally distinguish speech signals processed by different enhancement methods (as metadata), the model has a better understanding of differentiating the front-end processing of HA, potentially improving the prediction performance. In the following discussion, we refer to this classifier as the system classifier (SC). Third, MBI-Net+ utilizes the HASPI measure as a complementary metric, employing a multi-task learning criterion for model training. This method builds upon the previous study by \cite{zezario2023multitask}, which integrates additional related metrics correlated with the main metric to enhance prediction performance.

To implement MBI-Net+, the input speech data is initially divided into two channels, with one channel dedicated to the left ear and the other to the right ear of an HA. Each channel is processed by a unified module that combines feature extraction and neural network modeling. These two branches of modules simultaneously predict frame-level scores of intelligibility and HASPI, while also extracting embedding representations for the SC module. The predicted frame-level scores from both branch modules are then combined and fused via two task-specific modules to obtain final utterance-level scores for intelligibility and HASPI. Each task-specific module consists of a linear layer followed by a global average pooling layer. The embedding representations from each task-independent module are concatenated and processed by a dense layer with softmax to form the SC module that predicts the HA system.

Experimental results demonstrate that MBI-Net+ achieves a 7.11\% improvement in prediction performance, measured by root mean square error (RMSE), compared to the original MBI-Net model. This confirms the advantages of Whisper in deploying cross-domain features over the self-supervised learning (SSL) speech model, WavLM \cite{chen2021wavlm}, as well as the benefits of incorporating additional metadata. In the Clarity Prediction Challenge 2023 dataset, MBI-Net+ ranked third in overall performance compared to other non-intrusive systems, with comparable correlation values (0.76 (MBI-Net+) vs 0.78 (1st place \cite{cuervo2024speech}) vs 0.77 (2nd place \cite{mogridge2024nonintrusive})). It's noteworthy that MBI-Net+ requires less GPU memory compared to \cite{cuervo2024speech, mogridge2024nonintrusive}. This reduction can be attributed to MBI-Net+'s utilization of the last layer of Whisper to extract embedding representations. Unlike other approaches, MBI-Net+ doesn't require the extraction of each individual transformer output of Whisper and additional processing before forwarding it to the subsequent stage of training.

The remainder of this paper is organized as follows. Section II presents the proposed MBI-Net+. Section III describes the experimental setup and results. Finally, Section IV provides conclusions and future work.
%\graphicspath{ {./images/} }
%\begin{figure}[t]
%\centering
%\includegraphics[width=8cm]{Figure/MBI-Net_.pdf} 
%\caption{Architecture of the MBI-Net+ model.} 
%\label{fig:MOSA-Net}
%\end{figure}

\graphicspath{ {./images/} }
\begin{figure}[t]
\centering
\includegraphics[width=7.cm]{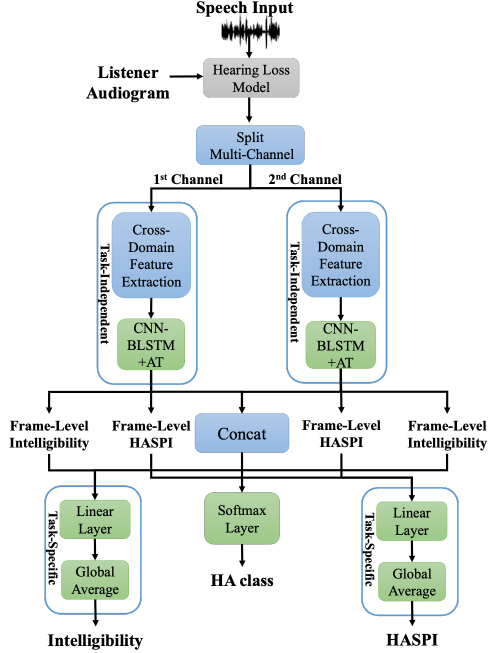} 
\caption{Architecture of the MBI-Net+ model.} 
\label{fig:MOSA-Net}
\end{figure}

\graphicspath{ {./images/} }
\begin{figure}[t]
\centering
\includegraphics[width=6cm]{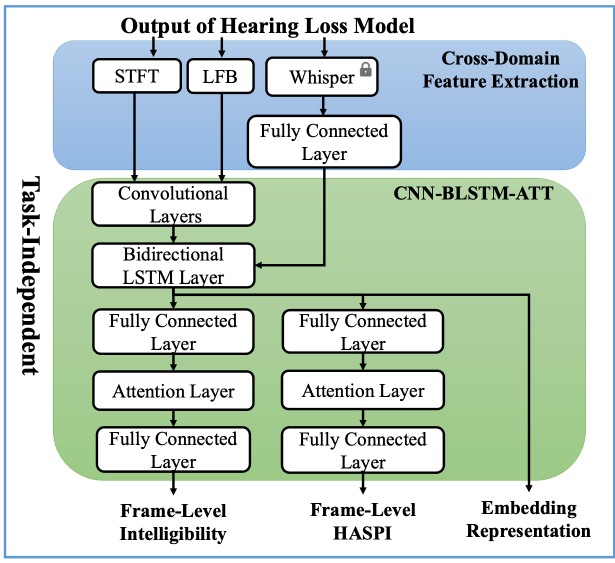} 
\caption{Illustration of extracting cross-domain features and estimating frame-level intelligibility scores using the CNN-BLSTM+AT architecture.} 
\label{fig:cross-domain}
\end{figure}

\section{MBI-Net+}

The proposed MBI-Net+ consists of multi-branched task-independent modules characterizing each speech signal channel in a binaural HA. Then, task-specific modules are used to predict intelligibility and HASPI scores, respectively. The overall architecture of MBI-Net+ and its task-independent module are illustrated in Figs. 1 and 2, respectively. Firstly, we utilize the pre-trained Whisper model instead of the pre-trained SSL models, HuBERT~\cite{hubert} and WavLM~\cite{chen2021wavlm}, employed in MBI-Net, to incorporate cross-domain features. We assume that Whisper features provide better phonetic information than SSL features, considering their larger training size and access to transcripts during deployment. Secondly, we include an SC module to categorize enhancement systems that were used to process speech signals. Considering their distinct characteristics, we assume that if MBI-Net+ can distinguish the different properties of processed speech signals, it can achieve higher prediction capabilities. This improvement is due to MBI-Net+ incorporating additional information when capturing acoustic information, potentially reducing overfitting during training. Thirdly, we introduce HASPI as a complementary metric. We assume HASPI correlates with subjective intelligibility score and, therefore, can potentially reduce overfitting and improve the prediction performance in multi-task learning scenarios \cite{zezario2023multitask}. Furthermore, The loss function, $O$, for training MBI-Net+ is as follows:
\begin{equation}
\label{eq:overall_loss}
   \small
    \begin{array}{c}
    O = \gamma_{1}\L_{Int} + \gamma_{2}\L_{HASPI} + \gamma_{3}\L_{CE},
    \end{array} 
\end{equation}
where $\L_{CE}$ represents the cross-entropy calculation from the SC module between the estimated and reference types of enhancement methods (front-end processors in HA systems). Specifically, we employ ten classes to account for the ten different enhancement systems used as input in our MBI-Net+. $\gamma_{1}$, $\gamma_{2}$, $\gamma_{2}$ are the weights between losses. $\L_{Int}$ is calculated as
\begin{equation}
\label{eq:int_loss}
   \small
    \begin{array}{c}
    \L_{Int}=\frac{1}{U}\sum\limits_{u=1}^U [(I_u-{\hat{I}}_u)^2+\frac{\alpha}{F_u}\sum\limits_{f=1}^{F_u}(I_u-{\hat{i}}_{u,f})^2] + \\ \L_{left-int} + \L_{right-int},
    \end{array} 
\end{equation}
where $I_u$, ${\hat{I}}_u$, and ${\hat{i}}_{u,f}$ denote the true utterance-level score, predicted utterance-level score, and predicted frame-level score (merged from the first and second channels by the linear layer in Figure 1) of intelligibility, respectively;
$U$ denotes the total number of training utterances;
$F_u$ denotes the number of frames in the $u$-th training utterance; $\alpha$ is a weight between utterance-level and frame-level losses; $\L_{left-int}$ and $\L_{right-int}$ are the frame-level losses of the left (\textit{l}) branch (i.e., the first channel) and right (\textit{r}) branch (i.e., the second channel) in frame-level intelligibility estimation (as shown in Figure 2), which are calculated as
\begin{equation}
\label{eq:int_lr_loss}
   \small
    \begin{array}{c}
    \L_{left-int}=\frac{\alpha^{l}}{F_u}\sum\limits_{f=1}^{F_u}(I_u-{\hat{i}}^{l}_{u,f})^2\\
    \L_{right-int}=\frac{\alpha^{r}}{F_u}\sum\limits_{f=1}^{F_u}(I_u-{\hat{i}}^{r}_{u,f})^2,\\
    \end{array} 
\end{equation}
where $\alpha^{l}$ and $\alpha^{r}$ are weights of \textit{l} and \textit{r} branches, respectively, and ${\hat{i}}^{l}_{u,f}$ and ${\hat{i}}^{r}_{u,f}$ denote the predicted frame-level scores of \textit{l} and \textit{r} branches, respectively.
$\L_{HASPI}$ is calculated as
\begin{equation}
\label{eq:haspi-loss}
   \small
    \begin{array}{c}
    \L_{HASPI}=\frac{1}{U}\sum\limits_{u=1}^U [(H_u-\hat{H}_u)^2+\frac{\beta}{F_u}\sum\limits_{f=1}^{F_u}(H_u-{\hat{h}}_{u,f})^2] + \\ \L_{left-haspi} + \L_{right-haspi},
    \\ 
    \end{array} 
\end{equation}
where $H_u$, $\hat{H}_u$, and ${\hat{h}}_{u,f}$ denote the true utterance-level score, predicted utterance-level score, and predicted frame-level score (merged from the first and second channels by the linear layer in Figure 1) of HASPI, respectively; $\beta$ is a weight between utterance-level and frame-level losses, and $\L_{left-haspi}$ and $\L_{right-haspi}$ are the frame-level losses of \textit{l} branch and \textit{r} branch in frame-level HASPI estimation calculated as 
\begin{equation}
\label{eq:haspi_lr_loss}
   \small
    \begin{array}{c}
    \L_{left-haspi}=\frac{\beta^{l}}{F_u}\sum\limits_{f=1}^{F_u}(H_u-{\hat{h}}^{l}_{u,f})^2\\
    \L_{right-haspi}=\frac{\beta^{r}}{F_u}\sum\limits_{f=1}^{F_u}(H_u-{\hat{h}}^{r}_{u,f})^2,\\
    \end{array} 
\end{equation}
where $\beta^{l}$ and $\beta^{r}$ are weights of \textit{l} and \textit{r} branches, and ${\hat{h}}^{l}_{u,f}$ and ${\hat{h}}^{r}_{u,f}$ denote the predicted frame-level scores of \textit{l} and \textit{r} branches, respectively. Furthermore, we hypothesize that the supplementary information from HASPI and the SC module can improve overall prediction performances. Subsequently, the corresponding frame-level scores are combined and integrated through two linear layers. This combination produces the final predictive scores for intelligibility and HASPI scores.

\section{Experiments}

\subsection{Experimental Setup}
The 2023 Clarity Prediction Challenge (CPC) dataset \cite{barker2024} comprises scenes associated with 6 talkers, 10 enhancement methods (accordingly 10 HA systems) from the 2022 Clarity Enhancement Challenge \cite{10094918}, and 25 listeners who rated the intelligibility scores. To elaborate, the dataset is divided into three tracks. The first track consists of 2779 utterances, the second track consists of 2796 utterances, and the third track consists of 2772 utterances. For each track, we selected 90\% of the utterances as the training set and the rest as the development set. Additionally, the three tracks have 305, 294, and 298 test utterances, respectively. Please note that the test involves unseen listeners and unseen HA systems. Specifically, the model is trained and tested on three tracks. In our experiment, the overall performance across all tracks is denoted as All. Three evaluation metrics, namely root mean square error (RMSE), linear correlation coefficient (LCC), and Spearman's rank correlation coefficient (SRCC) \cite{srcc} were used to evaluate the prediction performance. A lower RMSE value means that the predicted scores are closer to the ground-truth scores (lower is better). In contrast, higher LCC and SRCC values indicate a higher correlation between the predicted scores and the ground-truth scores (higher is better). All models compared in this paper were trained and evaluated on the CPC 2023 dataset.

\graphicspath{ {./images/} }
\begin{figure}[t]
\centering
\includegraphics[width=5.8 cm]{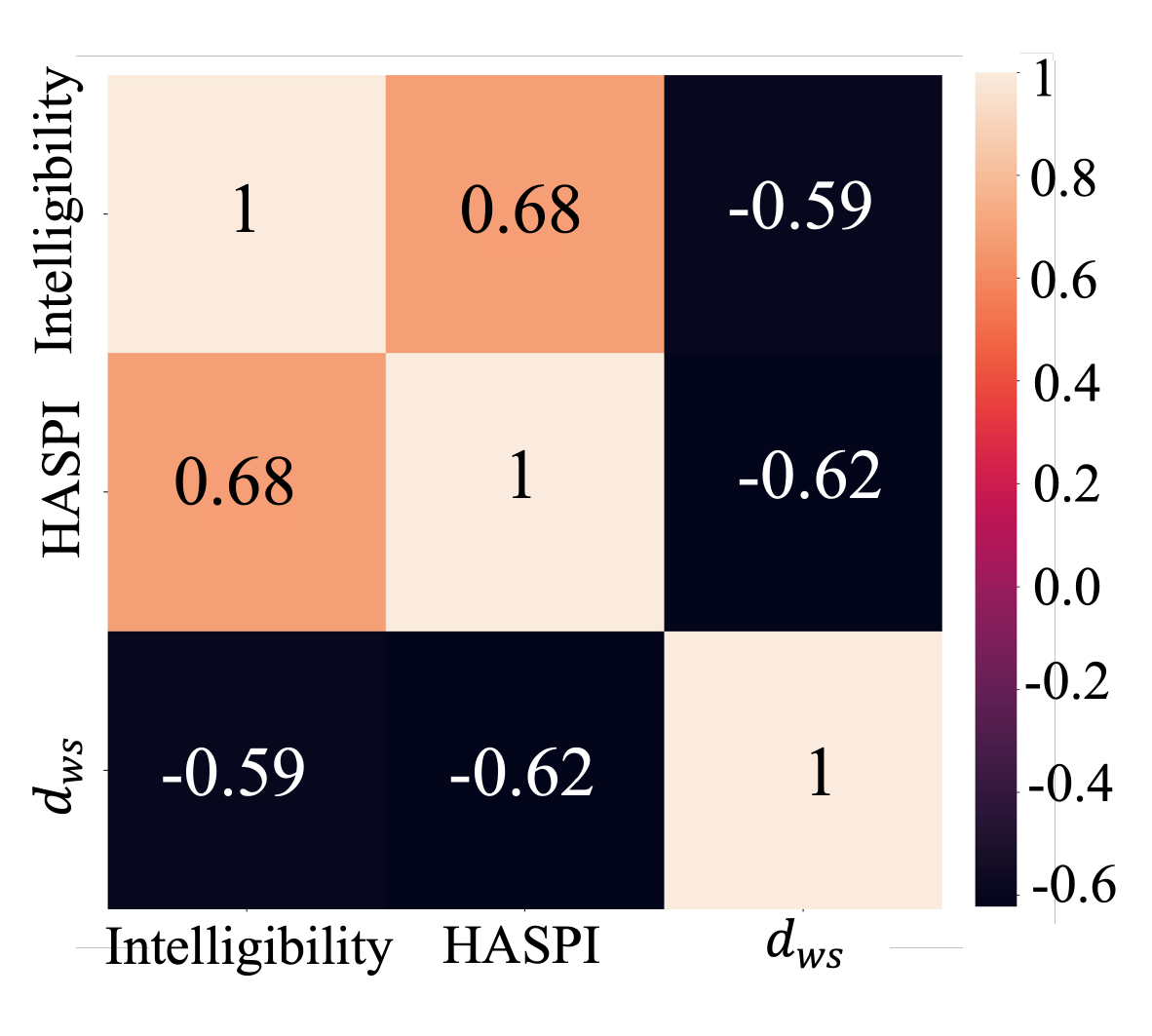} 
\caption{Correlation analysis between Intelligibility, HASPI, and Whisper.} 
\label{fig:corr}
\end{figure}

\begin{table}[t]
\caption{LCC, SRCC, and MSE results of MBI-Net, MBI-Net+(\textit{w/o SC}) and MBI-Net+ on the development set.}
\footnotesize
\begin{center}
 \begin{tabular}{c||c||c||c||c} 
 \hline
 \hline
 \textbf{Model} &\textbf{Feature}&\textbf{LCC} & \textbf{SRCC} & \textbf{RMSE}  \\ [0.5ex] \cline{2-5}
 \hline\hline
  \multicolumn{5}{c} {Track 1}
\\ \hline
MBI-Net \cite{edozezario22_interspeech}&WavLM&0.724&0.719&28.707\\\hline
MBI-Net+(\textit{w/o SC-H}) &Whisper&0.754&0.738&27.221\\\hline
MBI-Net+(\textit{w/o SC}) &Whisper&0.758&0.748&26.857\\\hline\
MBI-Net+ &Whisper&\textbf{0.773}&\textbf{0.763}&\textbf{26.081}\\\hline
 \hline
  \multicolumn{5}{c} {Track 2} \\
 \hline
MBI-Net \cite{edozezario22_interspeech}&WavLM&0.742&0.749&29.328\\\hline
MBI-Net+(\textit{w/o SC-H}) &Whisper&0.794&0.788&26.011\\\hline
MBI-Net+(\textit{w/o SC}) &Whisper&0.799&0.797&25.844\\\hline
MBI-Net+ &Whisper&\textbf{0.804}&\textbf{0.799}&\textbf{25.392}\\\hline
 \hline
  \multicolumn{5}{c} {Track 3} \\
 \hline
MBI-Net \cite{edozezario22_interspeech}&WavLM&0.795&0.772&27.155\\\hline
MBI-Net+(\textit{w/o SC-H}) &Whisper&0.797&0.764&24.706\\\hline
MBI-Net+(\textit{w/o SC}) 
&Whisper&0.801&0.765&24.800\\\hline
MBI-Net+ &Whisper&\textbf{0.817}&\textbf{0.773}&\textbf{23.575}\\\hline
 \hline
\end{tabular}
\end{center}
\end{table}

\subsection{Correlation of Intelligibility, HASPI, and Distance of Whisper Embeddings}
In the first experiment, we aim to analyze the correlation between subjective intelligibility, HASPI, and the distance of Whisper embeddings, $d_{ws}$. A higher correlation indicates a stronger relationship. We aim to investigate whether considering these attributes, MBI-Net+ can achieve better predictions. We estimate $d_{ws}$ using the following equations

%\begin{equation}
%\label{eq:loss4}
%    \begin{array}{c}
%\textbf{Y}_{ws}=E_{ws}(F_{ws}(\textbf{Y}))
%\\
%\textbf{X}_{ws}=E_{ws}(F_{ws}(\textbf{X}))
%    \end{array} 
%\end{equation}

\begin{equation}
\label{eq:loss5}
%d_{ws}=\sum\limits_{t,f}^{T,F}(E_{ws}(F_{ws}(\textbf{X}[t,f]))-E_{ws}(F_{ws}(\textbf{Y})[t,f]))^2
d_{ws}=\sum\limits_{t,f}^{T,F}(\textbf{X}_{ws}[t,f]-\textbf{Y}_{ws}[t,f])^2
\end{equation}
where $\textbf{X}_{ws}[t,f]$ and $\textbf{Y}_{ws}[t,f]$ represent the last encoder output of Whisper at the $t$-th time index and $f$-th element from the clean waveform $\textbf{X}$ and enhanced waveform $\textbf{Y}$, respectively. \textit{T} and \textit{F} denote the time length and the feature dimension of the embedding representation.

Figure 3 illustrates a moderate correlation score of 0.68 between HASPI and intelligibility, suggesting HASPI's ability to surrogate subjective intelligibility. Interestingly, the distance of Whisper embedding, namely $d_{ws}$, demonstrates correlation scores of -0.59 and -0.62 with intelligibility and HASPI, respectively. This suggests a moderate degree of correlation between Whisper's representation and both intelligibility and HASPI. Therefore, we assume that employing Whisper embedding could be beneficial for estimating subjective intelligibility and HASPI scores.
%These findings prompt further investigation into the potential effectiveness of MBI-Net+ by incorporating HASPI and Whisper as additional attributes.

\begin{table}[t]
\caption{LCC, SRCC, and MSE results of MBI-Net, MBI-Net+(\textit{w/o SC}), MBI-Net+(\textit{w/o SC-H}), and MBI-Net+ on the test set.}
\footnotesize
\begin{center}
 \begin{tabular}{c||c||c||c||c} 
 \hline
 \hline
 \textbf{Model} &\textbf{Feature}&\textbf{LCC} & \textbf{SRCC} & \textbf{RMSE}  \\ [0.5ex] \cline{2-5}
 \hline\hline
  \multicolumn{5}{c} {Track 1}
\\ \hline
MBI-Net \cite{edozezario22_interspeech}&WavLM&0.669&0.665&30.260\\\hline
MBI-Net+(\textit{w/o SC-H}) &Whisper&0.703&0.690&29.270\\\hline
MBI-Net+(\textit{w/o SC}) &Whisper&0.711
&0.692&29.010\\\hline
MBI-Net+ &Whisper&\textbf{0.721}&\textbf{0.714}&\textbf{28.370}\\\hline
 \hline
  \multicolumn{5}{c} {Track 2} \\
 \hline
MBI-Net \cite{edozezario22_interspeech}&WavLM&0.710&0.710&29.660\\\hline
MBI-Net+(\textit{w/o SC-H}) &Whisper&0.751&0.737&26.730\\\hline
MBI-Net+(\textit{w/o SC}) &Whisper&\textbf{0.765}&\textbf{0.764}&26.440\\\hline
MBI-Net+ &Whisper&0.754&0.737&\textbf{25.920}\\\hline
 \hline
  \multicolumn{5}{c} {Track 3} \\
 \hline
MBI-Net \cite{edozezario22_interspeech}&WavLM&0.810&0.834&23.920\\\hline
MBI-Net+(\textit{w/o SC-H}) &Whisper&0.808&\textbf{0.840}&24.040\\\hline
MBI-Net+(\textit{w/o SC})&Whisper&\textbf{0.826}&0.834&\textbf{23.340}\\\hline
MBI-Net+ &Whisper&0.813&0.814&23.740\\\hline
 \hline
  \multicolumn{5}{c} {Track All} \\
 \hline
MBI-Net \cite{edozezario22_interspeech}&WavLM&0.724&0.729&28.100\\\hline
MBI-Net+(\textit{w/o SC-H}) &Whisper&0.754&0.752&26.790\\\hline
MBI-Net+(\textit{w/o SC})&Whisper&\textbf{0.767}&0.765&26.390\\\hline
MBI-Net+ &Whisper&0.764&\textbf{0.767}&\textbf{26.100}\\\hline
 \hline
\end{tabular}
\end{center}
\end{table}

\begin{table}[t]
\caption{RMSE and LCC scores of all compared systems from the First and Second Clarity Challenge on the test set.}
\footnotesize
\begin{center}
 \begin{tabular}{c||c||c||c} 
 \hline
 \hline
 \textbf{System} & \textbf{Non-Intrusive} & \textbf{RMSE} & \textbf{LCC}  \\ [0.5ex] 
 \hline
\hline
E011 \cite{cuervo2024speech}&Yes&\textbf{25.1}&\textbf{0.78} \\ \hline 
E002 \cite{mogridge2024nonintrusive}&Yes&25.3&0.77 \\ \hline 
MBI-Net+&Yes&26.1&0.76 \\ \hline
MBI-Net+(\textit{w/o SC})&Yes&26.4&0.76 \\ \hline
MBI-Net+(\textit{w/o SC-H})&Yes&26.8&0.75 \\ \hline
%MBI-Net+&Yes&26.8&0.75 \\ \hline
%MBI-Net+&Yes&26.8&0.754 \\ \hline
E025 \cite{barker2024} &Yes&27.9&0.72 \\ \hline 
MBI-Net \cite{edozezario22_interspeech}&Yes&28.1&0.72 \\ \hline
Baseline \cite{barker2024}&No&28.7&0.70 \\ \hline 
E003 \cite{barker2024} &Yes&31.1&0.64 \\ \hline 
E024 \cite{barker2024} &Yes&31.7&0.62 \\ \hline 
E015 \cite{barker2024} &Yes&35.0&0.60 \\ \hline 
E020 \cite{barker2024} &Yes&39.8&0.33 \\ \hline 
Prior \cite{barker22_interspeech}&No&40 & - \\ \hline 
 \hline

\end{tabular}
\end{center}
\end{table}

\subsection{Comparing MBI-Net+ with Original MBI-Net}

In the first experiment, we aim to compare the performance of MBI-Net+ and MBI-Net. The MBI-Net+ entails: 1) employing a 257-dimensional spectral feature; 2) leveraging Whisper medium as the pre-trained model, with the final layer's output serving as input for our MBI-Net+; 3) utilizing four convolutional layers each comprising channels with sizes of 16, 32, 64, and 128, one-layered Bidirectional Long Short-Term Memory (BLSTM) with 128 nodes, fully connected layer containing 128 neurons, attention mechanism, and one neuron of frame level score; 5) leveraging two task-specific modules and SC module (with ten classes); 6) setting a learning rate of 0.0001 with Adam optimizer \cite{KingmaB14}. Note that we developed three different versions of MBI+, namely MBI-Net+(\textit{w/o SC}), MBI-Net+(\textit{w/o SC-H}), and MBI-Net+. MBI-Net+ represents the complete configuration in Figure 1, and MBI-Net+(\textit{w/o SC}) follows the same model architecture as shown in Figure 1, except not including the SC module. On the other hand, MBI-Net+(\textit{w/o SC-H}) does not include HASPI and SC modules.

As shown in Tables 1 and 2, all versions of MBI-Net+ can outperform MBI-Net. Please note that MBI-Net+(\textit{w/o SC-H}) shares the same model architecture as MBI-Net but uses different speech representations (SSL for MBI-Net and Whisper for MBI-Net+(\textit{w/o SC-H})). From Tables 1 and 2, the comparison between MBI-Net+(\textit{w/o SC-H}) and MBI-Net first confirms the advantage of incorporating Whisper embeddings for deploying cross-domain features. Next, the comparison between MBI-Net+(\textit{w/o SC}) and MBI-Net+(\textit{w/o SC-H}) confirms the effectiveness of utilizing supplementary HASPI metric which has moderate correlation with  the main objective metric in preventing overfitting in multi-task learning scenario. Finally, the comparison between MBI-Net+ and MBI-Net+(\textit{w/o SC}) shows that the SC module contributes to better prediction performance, confirming the advantages of incorporating additional metadata of the speech signals.

\subsection{Comparison with Other Prediction Models for HA}
In the third experiment, we compare MBI-Net+ models with other speech intelligibility prediction models for HA. Four systems (E011 \cite{cuervo2024speech}, E002 \cite{mogridge2024nonintrusive}, MBI-Net \cite{edozezario22_interspeech}, E025 \cite{barker2024}) utilized acoustic features from large pre-trained models, while the other four \cite{barker2024} (E003, E024, E015, E020) utilized signal processing techniques for feature extraction. %Moreover, all the systems employed an ensemble learning model. 
The overall RMSE and LCC scores from the three tracks are shown in Table 3. We first confirm that all versions of MBI-Net+ can perform better than the baseline \cite{barker2024} model that adopts an intrusive-based method. 
Furthermore, our models achieve good performance, ranking third (MBI-Net+) overall among non-intrusive speech intelligibility prediction models. The LCC score of MBI-Net+ is 0.76, which is very close to the best score (0.78) and the second-best score (0.77). It's noteworthy that MBI-Net+ requires less GPU memory compared to \cite{cuervo2024speech, mogridge2024nonintrusive}. Considering MBI-Net+ doesn't require the extraction of each individual transformer output of Whisper and additional processing before forwarding it to the subsequent stage of training. Finally, this result demonstrates the potential benefits of adopting a multi-task, multi-branched model architecture and incorporating Whisper-based cross-domain features along with adding supplementary metrics and the SC module into better prediction capabilities.

\section{Conclusion}
In this study, we have proposed MBI-Net+, a novel speech intelligibility prediction model for HA. MBI-Net+ implements a multi-branched architecture that incorporates Whisper-based cross-domain features and includes the supplementary HASPI metric and the SC module. Experimental results first confirm a moderate correlation between subjective intelligibility, HASPI, and distance of Whisper embeddings, $d_{ws}$, suggesting potential advantages for their joint combinations. Next, we confirm the effectiveness of MBI-Net+ by achieving a notable improvement over MBI-Net and yielding a top-3 performance among non-intrusive systems in the Clarity Prediction Challenge 2023. Since MBI-Net+ only utilizes the final layer of the pre-trained Whisper model, this approach avoids the heavy computational requirements of other top-performing systems. In future work, we plan to investigate the incorporation of other types of metadata from speech signals and model architectures when deploying speech intelligibility prediction models for HA.

\bibliographystyle{IEEEtran}
\bibliography{mybib}

% \begin{thebibliography}{9}
% \bibitem[1]{Davis80-COP}
%   S.\ B.\ Davis and P.\ Mermelstein,
%   ``Comparison of parametric representation for monosyllabic word recognition in continuously spoken sentences,''
%   \textit{IEEE Transactions on Acoustics, Speech and Signal Processing}, vol.~28, no.~4, pp.~357--366, 1980.
% \bibitem[2]{Rabiner89-ATO}
%   L.\ R.\ Rabiner,
%   ``A tutorial on hidden Markov models and selected applications in speech recognition,''
%   \textit{Proceedings of the IEEE}, vol.~77, no.~2, pp.~257-286, 1989.
% \bibitem[3]{Hastie09-TEO}
%   T.\ Hastie, R.\ Tibshirani, and J.\ Friedman,
%   \textit{The Elements of Statistical Learning -- Data Mining, Inference, and Prediction}.
%   New York: Springer, 2009.
% \bibitem[4]{YourName17-XXX}
%   F.\ Lastname1, F.\ Lastname2, and F.\ Lastname3,
%   ``Title of your INTERSPEECH 2021 publication,''
%   in \textit{Interspeech 2021 -- 20\textsuperscript{th} Annual Conference of the International Speech Communication Association, September 15-19, Graz, Austria, Proceedings, Proceedings}, 2020, pp.~100--104.
% \end{thebibliography}

\end{document}